# SIMULATION OF THE ROTATIONAL SUBLEVELS OF THE GROUND AND THE FIRST EXCITED VIBRATIONAL STATES OF $H_2S$ USING RESUMMATION METHODS

O.V. Naumenko, E. R. Polovtseva, A.D. Bykov

*V.E. Zuev Institute of Atmospheric Optics, SB RAS, 1, Academician Zuev Square, Tomsk 634021, Russia*

## ABSTRACT

The rotational sublevels of the key (000) and (010) vibrational states of the $H_2S$ molecule were modeled with an accuracy close to experimental uncertainty using the generating function and Euler approaches. The predictive ability of the Hamiltonian's parameters derived is tested against variational calculations. Comparison of transitions wavenumbers obtained from the presently calculated set of the $H_2S$ (000) and (010) energy levels with simulated (000)-(000), (010)-(010) transitions included in HITRAN 2916 database revealed a large discrepancy up to 44 $cm^{-1}$. Large sets of accurate rotational sublevels of the (000) and (010) states are calculated.

## I. INTRODUCTION

The ground (000) and first excited (010) are the key vibrational states of the $H_2S$ molecule, as majority of the observed rotational - vibrational transitions in absorption originate from the rotational sublevels of these states. The experimental energy levels of the lower vibrational state are preferably used to determine the upper level of transition. However, the experimental rotational energy levels of the $H_2S$ ground vibrational state were not available until recently [1]. In Ref. [1] a set of consistent experimental energy levels was derived from simultaneous consideration of the observed rovibrational transitions of $H_2S$ molecule. The accuracy of the energy level determination in [1] depends on the accuracy of transitions involving this level, and in case of a lack of input data or their inconsistency, the resulted level may be considerably distorted. On the other hand, the set of experimental energy levels [1] is limited in $J$ and $K_a$ rotational quantum numbers. Then, an accurate simulation could support the energy levels values [1], or even improve them by rejecting less accurate transition involved in level determination. In this contribution an accurate set of calculated (000) and (010) rotational energy levels is obtained for the $H_2S$ molecule from the fitting to the observed energy levels [1] using the effective rotational Hamiltonian based on different resummation methods of the divergent perturbation series.

## II. EFFECTIVE HAMILTONIAN AND ENERGY LEVELS FITTING

The rotational Hamiltonian written through the generating function $G$ [2] has the form:

$$H_{rot}^{G} = \sum_{n,m} g_{nm} J^{2n} \{G(\alpha^{(J)})\}^{m} + \sum_{n,m} u_{nm} J^{2n} [(J_{+}^{2} + J_{-}^{2}), \{G(\alpha^{(J)})\}^{m}]_{+}, \qquad (1)$$

Where the generating function is defined according to [3]

$$G(\alpha^{(J)}) = (2/\alpha^{(J)}) \left\{ \sqrt{1+\alpha^{(J)} J_{z}^{2}} - 1 \right\}.$$

The $J$-dependence of $\alpha^{(J)}$ in the generating function is given by the development

$$\alpha^{(J)} = \sum_{n} \alpha_{n} J^{2n}.$$

Numerical parameters $g_{nm}$, $u_{nm}$, and $\alpha_n$ of Hamiltonian (1) are adjustable spectroscopic parameters. This representation of the rotational operator doesn't contain divergent series essential for the conventional perturbative approach, and proved to be much more accurate not only in the fitting, but also in the extrapolation calculations. Relations of the $H_{rot}^{G}$ parameters with Watson's constants are discussed in [4]. Spectroscopic parameters of Hamiltonian (1) were determined in a least square fit to experimental rotational energies of each vibrational state.

Experimental energy levels derived in recent paper [1] for the (000) and (010) state of the $H_2S$ molecule were introduced into the least squares fitting using the effective Hamiltonian (1). The resulting parameters with 1 sigma confidential intervals are shown in Table 1. An RMS deviations on the order of 0.0002 cm$^{-1}$ what is close to experimental accuracy were obtained by varying of 37 and 35 parameters of Hamiltonian (1) for the (000) and (010) states, respectively.

In fact, we used Watson Hamiltonian, Pade-Borel approximants, generating functions (GF), and Euler resummation ( Ref. [6]) in the energy levels calculation. It seems that the GF approach provides most accurate fitting and extrapolation, that is why we show the illustrations which, mainly, concern only this method. To compare our results with the literature, we calculate large sets of the 000 and 010 energy levels and constructed pure rotational (000)-(000) and (010)-(010) transitions. Then these transitions were compared with the HITRAN 2016 data.

### III. COMPARISON WITH THE HITRAN 2016 DATABASE AND VARIATIONAL CALCULATIONS

HITRAN 2016 database contains both pure experimental and simulated pure rotational transition of $H_2S$ between 0 and 610 cm$^{-1}$. These transitions are taken, mostly, from Ref. [5], where the experimental $H_2S$ spectra between 40 and 360 cm$^{-1}$ were assigned and modeled using Pickett program suite [6] which offers also the Euler resummation method. In the fitting in [5] not only new experimental data were used, but all available information from the literature was included. The resulting list includes together with experimental transitions a large number of simulated frequencies, mostly for weak lines, with estimated accuracy on positions up to 0.033 cm$^{-1}$.

The pure rotational (000)-(000) and (010)-(010) transitions of H$_2$S between 45 and 360 cm$^{-1}$ included into HITRAN data base were compared with those calculated based on parameters derived in this study. The comparison is included into Figs. 1-2 and Tables 2-3. It seen from Figs. **1 and 2** that there are large (up to 44 cm$^{-1}$) disagreements between our and Ref. **[5]** data.

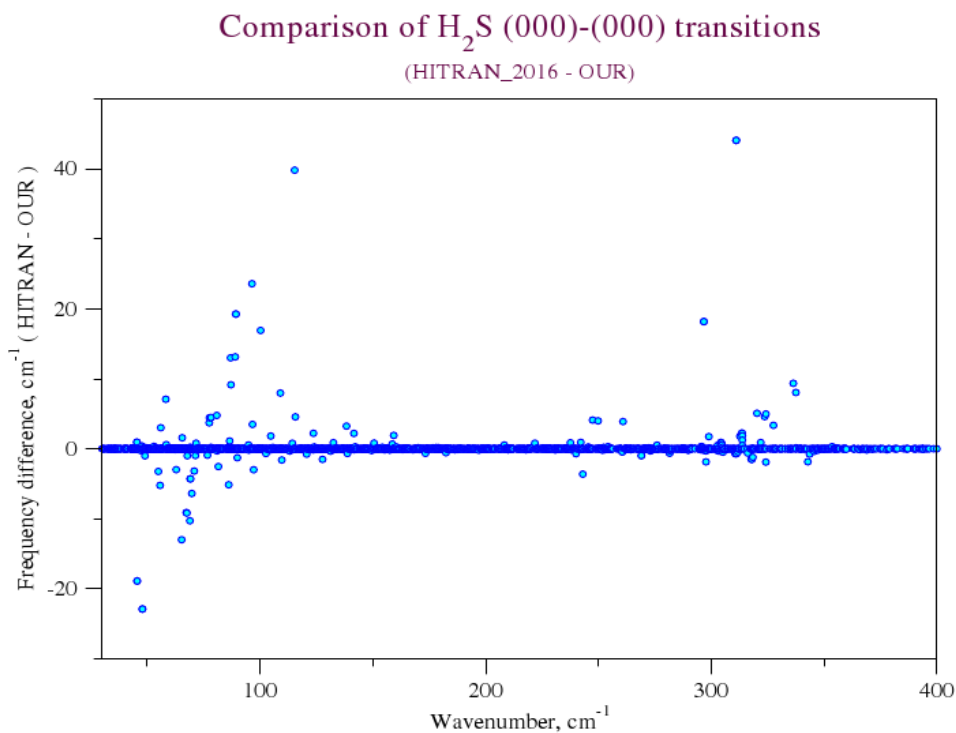

**Fig.1.** Comparison of the (000)-(000) pure rotational transitions of H$_2$S molecule included into HITRAN 2016 database and those calculated from the parameters presented in Table 1.

To further validate our H$_2$S calculations based on the Generation function approach, the pure rotational (010)-(010) transitions calculated in this paper were compared with recent variational calculation [7] **(see Fig.3)**. The comparison was limited to H$_2$S transitions from HITRAN 2016. It is obvious from the figure that our simulation agrees much better with variational calculation than with simulation from Ref. [5] with maximal deviation not exceeding 0.4 cm$^{-1}$.

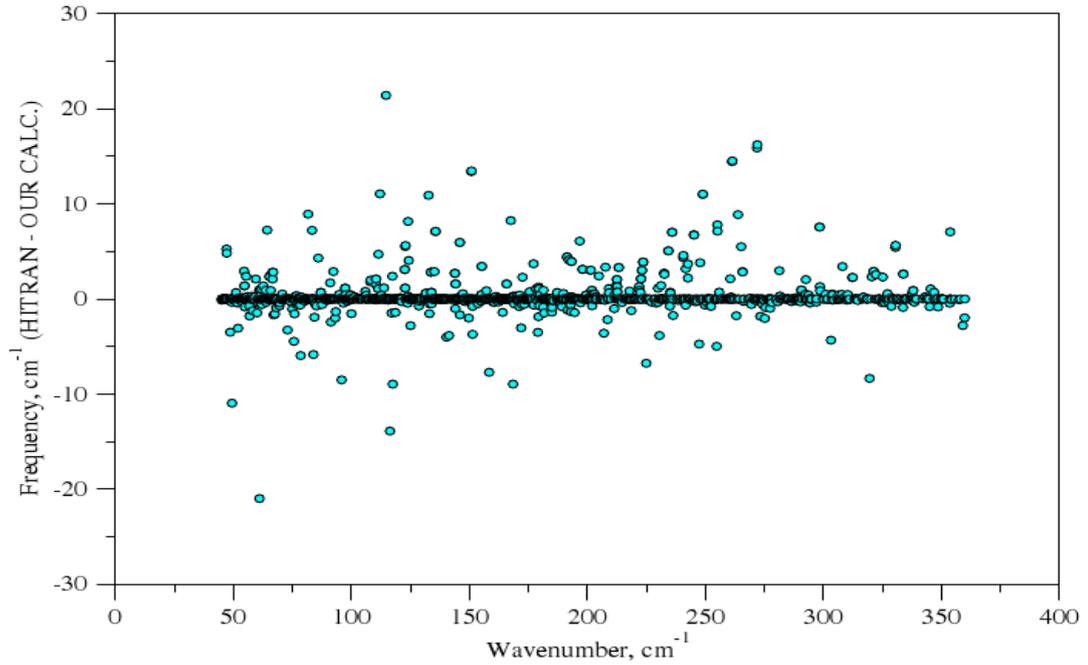

**Fig.2.** Comparison of the (010)-(010) pure rotational transitions of H$_2$S molecule included into HITRAN 2016 database and those calculated from the parameters presented in Table 1.

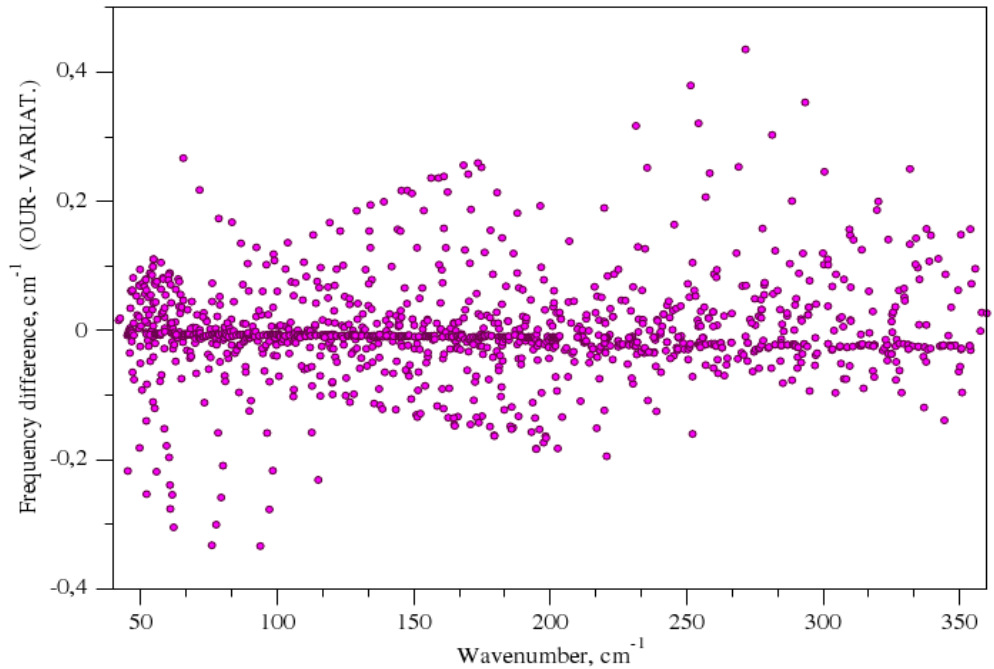

**Fig. 3.** Comparison of the (010)-(010) pure rotational transitions of H$_2$S calculated in this paper with recent variational calculation [7].

# IV. MODELING AND PREDICTING THE H$_2$S TRANSITIONS USING EULER SUMMATION METHOD

A set of pure rotational (000)-(000), (010) – (010) H$_2$S transitions between 49 and 359 cm$^{-1}$ included in HITRAN 2016 database originate from the analysis of the terahertz H$_2$S spectrum undertaken in Ref.[5]. In this study newly recorded IR transitions were combined with the MW literature experimental data (1408 in total) and fitted using SPFIT/SPCAT computer code elaborated by H. Pickett [6]. Spectroscopic parameters obtained from the fitting were then used to predict a great number of weak H$_2$S transitions missing in experiment. Altogether, 2374 pure rotational transitions in the (000) and (010) vibrational states of H$_2$S, mostly weak, were predicted in the considered spectral region, while only 1160 experimental lines were initially assigned.

As it was illustrated in Section III, our calculation of the H$_2$S (000) and (010) rotational sublevels considerably disagree with Ref. [5] prediction: of 2374 compared transitions 789 have disagreement in positions between 0.05 and 44 cm$^{-1}$ (see also Tables 2-3). To clarify this situation, we made our own fitting of the (000)-(000) experimental H$_2$S transitions using SPFIT/SPCAT computer code [6]. The fitting was performed within the Euler method with an RMS of 0.00009 cm$^{-1}$ obtained for 1396 transitions by varying 39 parameters. Then the parameters obtained were used to predict the same set of weak transitions as in Ref. [5]. Comparison of the set of extrapolated transitions based on GF and Euler approaches is included in Fig. 4. Wavenumbers of both sets of transitions agree within RMS of 0.08 cm$^{-1}$, with maximum deviation not exceeding 0.7 cm$^{-1}$ what is very inspiring provided that a long extrapolation on J and $K_a$ rotational quantum numbers was used.

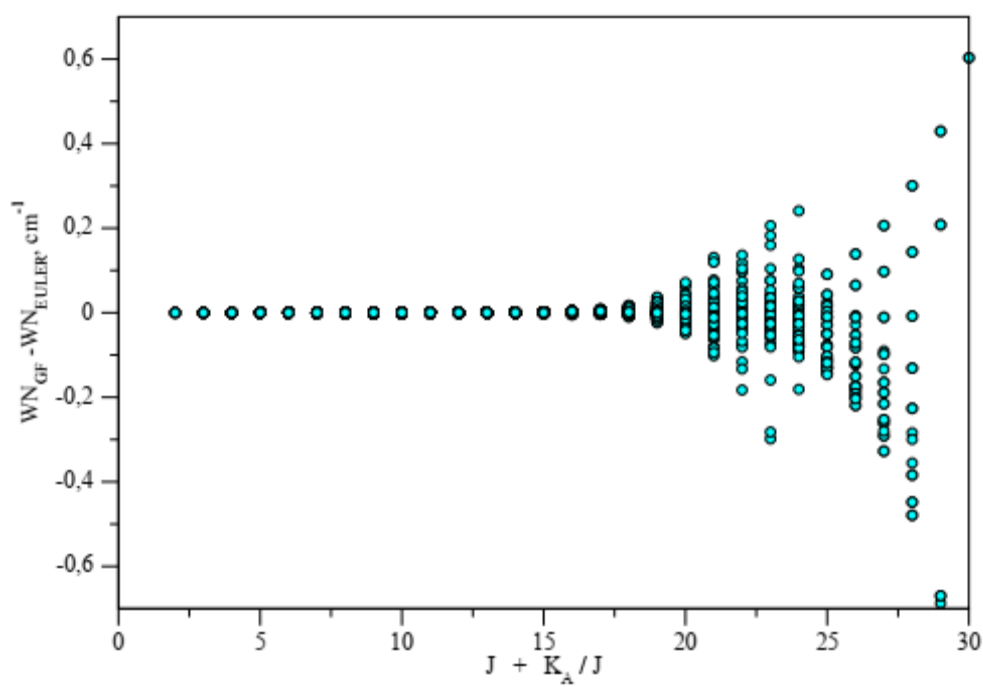

**Fig. 4.** Comparison of the (000)-(000) pure rotational transitions of H$_2$S calculated in this paper using Generation functions approach [2,3] and Euler method [6]. Comparison is limited to HITRAN 2016 data set.

## IV. CONCLUSION

Recently derived [1] experimental energy levels of the key (000) and (010) vibrational states of $H_2S$ molecule were modeled with the accuracy close to experimental one using the generating function and Euler methods [3,6]. The spectroscopic parameters obtained were used for extrapolating calculations of the (000) and (010) highly excited rotational energy levels and transition wavenumbers. These simulations were compared with the pure rotational (000)-(000) and (010) - (010) transitions from the HITRAN 2016 database based on the original data [5] revealing large - up to 44 cm$^{-1}$ disagreements.

Comparison of the presently calculated energy levels with recent accurate variational calculations [7] confirmed a high quality of simulation based on GF and Euler methods with maximal deviation not exceeding 0.4 and 0.76 cm$^{-1}$, respectively. A detailed consideration of the fitting procedure used in [5] allowed us to conclude that, in fact, the Watson - type rotational Hamiltonian with much worse predictive ability was used in [5] instead of Euler approach available in [6], as the main "a" and "b" Euler parameters were set to zero. In addition, it seems also that 12 blended (or misassigned) less accurate transitions participated in the fitting in [5], what resulted in increase of the number of varied parameters (46 compared to 39 in our study) and further degrading of the prediction accuracy.

Finally, large sets of accurate energy levels for the key (000) and (010) states of $H_2S$ molecule were generated on the base of the GF spectroscopic parameters. These energy levels can be widely used for the $H_2S$ spectra assignment and extrapolation. These data are available from the authors on request.

**Acknowledgment**

The work of Naumenko O.V. was supported by the Ministry of Science and Higher Education of the Russian Federation. The study of Polovtseva E.R. and Bykov A.D. was carried out with the financial support of the Russian Foundation for Basic Research (Grant No. 18-02-00462).

**Table 1.** Parameters of the effective Hamiltonian written through GF for the (000) and (010) vibrational states of the H$_2$S molecule.

| PAR | VALUE, cm$^{-1}$ (000) | (010) | PAR | VALUE, cm$^{-1}$ (000) | (010) |
|---|---|---|---|---|---|
| a1 | 0.27030(240)E-02 | 0.4832724(900)E-02 | g23 | | -0.3223(750)E-10 |
| a2 | 0.42914(390)E-05 | -0.4665(200)E-05 | g33 | 0.4339(210)E-14 | 0.2785(200)E-13 |
| a3 | -0.516(130)E-09 | 0.1350(160)E-07 | g04 | -0.37794(290)E-08 | -0.7541(520)E-08 |
| a4 | -0.11359(810)E-11 | -0.1595(610)E-10 | g24 | -0.15981(760)E-13 | -0.5134(780)E-13 |
| a5 | | 0.6212(700)E-14 | g15 | 0.3262(110)E-13 | 0.1073(100)E-12 |
| E | 0.0 | 1182.57696000 | g06 | -0.25538(650)E-13 | -0.7037(520)E-13 |
| g10 | 6.87446394(380) | 6.94661604(510) | (B-C)/4 | 1.07184244(250) | 1.13887827(250) |
| g20 | -0.6528257(890)E-03 | -0.75491(110)E-03 | u10 | -0.2956345(520)E-03 | -0.3466765(560)E-03 |
| g30 | 0.274666(780)E-06 | 0.36387(760)E-06 | u20 | 0.137178(420)E-06 | 0.181231(390)E-06 |
| g40 | -0.16808(280)E-09 | -0.1881(170)E-09 | u30 | -0.8453(140)E-10 | -0.93403(860)E-10 |
| g50 | 0.7714(350)E-13 | | u40 | 0.3906(180)E-13 | |
| g01 | 3.4856935(150) | 3.775483(170) | u01 | 0.132248(150)E-03 | 0.20853(210)E-04 |
| g11 | 0.2279979(450)E-02 | 0.273504(530)E-02 | u11 | -0.48015(170)E-06 | -0.66937(320)E-06 |
| g21 | -0.153536(490)E-05 | -0.19076(760)E-05 | u21 | 0.44811(760)E-09 | 0.8589(180)E-09 |
| g31 | 0.13757(210)E-08 | 0.2144(410)E-08 | u31 | -0.2849(110)E-12 | -0.9692(350)E-12 |
| g41 | -0.8121(310)E-12 | -0.1653(770)E-11 | u02 | 0.132058(260)E-05 | 0.217687(360)E-05 |
| g02 | -0.13471(210)E-02 | | u12 | -0.6565(130)E-09 | -0.13641(320)E-08 |
| g12 | 0.65431(330)E-05 | | u22 | 0.7936(230)E-12 | 0.34275(820)E-11 |
| g22 | -0.29608(910)E-08 | | u13 | -0.17641(350)E-11 | -0.9428(170)E-11 |
| g03 | -0.36253(420)E-05 | -0.8370(120)E-05 | u04 | 0.26969(360)E-11 | 0.6965(170)E-11 |
| g13 | | 0.2592(180)E-07 | | | |
| 37 param, 426 levels, RMS=0.000256 cm$^{-1}$ (000) | | | 35 param, 301 levels, RMS=0.00020 cm$^{-1}$ (010) | | |

**Table 2.** Comparison of the presently calculated using GF approach (000)-(000) pure rotational transition wavenumbers with those from HITRAN 2016 database for $H_2^{32}S$ molecule. [*]

| Wn, cm$^{-1}$ This work | Intensity Cm/mol | Wn, cm$^{-1}$ HIT - TW | E$_{lower}$, cm$^{-1}$ | VIB | J K$_a$ K$_c$ Upper | VIB | J K$_a$ K$_c$ Lower |
|---|---|---|---|---|---|---|---|
| 50.02800 | 1.326E-28 | -1.018666 | 4260.7181 | 000 | 22 14 8 | 000 | 22 13 9 |
| 51.16310 | 7.255E-29 | 7.083400 | 4652.9225 | 000 | 23 15 8 | 000 | 23 14 9 |
| 53.01100 | 1.133E-28 | 3.007489 | 4310.7461 | 000 | 22 15 7 | 000 | 22 14 8 |
| 58.18030 | 9.930E-29 | -3.247069 | 4594.7422 | 000 | 23 14 9 | 000 | 23 13 10 |
| 60.95640 | 5.766E-29 | -5.247698 | 4704.0856 | 000 | 23 16 7 | 000 | 23 15 8 |
| 63.94320 | 2.490E-28 | 1.560563 | 4196.7749 | 000 | 22 13 9 | 000 | 22 12 10 |
| 64.48180 | 1.004E-28 | -18.926174 | 4324.0917 | 000 | 21 20 1 | 000 | 21 19 2 |
| 64.48290 | 3.346E-29 | -18.926605 | 4324.0905 | 000 | 21 20 2 | 000 | 21 19 3 |
| 65.86190 | 1.005E-28 | -2.990330 | 4363.7571 | 000 | 22 16 6 | 000 | 22 15 7 |
| 68.76130 | 1.363E-27 | -1.001669 | 4038.7235 | 000 | 21 16 5 | 000 | 21 15 6 |
| 70.03120 | 3.487E-28 | 19.263254 | 4254.0605 | 000 | 21 19 2 | 000 | 21 18 3 |
| 70.04220 | 1.162E-28 | 19.263993 | 4254.0483 | 000 | 21 19 3 | 000 | 21 18 4 |
| 70.81840 | 9.997E-30 | -22.930724 | 4655.5386 | 000 | 22 20 2 | 000 | 22 19 3 |
| 70.82360 | 2.999E-29 | -22.934187 | 4655.5328 | 000 | 22 20 3 | 000 | 22 19 4 |
| 72.85110 | 6.345E-29 | 23.610855 | 4765.0420 | 000 | 23 17 6 | 000 | 23 16 7 |
| 73.18720 | 9.941E-28 | 4.439430 | 4107.4848 | 000 | 21 17 4 | 000 | 21 16 5 |
| 73.38850 | 5.487E-28 | -4.304845 | 4180.6720 | 000 | 21 18 3 | 000 | 21 17 4 |
| 73.48100 | 1.833E-28 | -4.310301 | 4180.5673 | 000 | 21 18 4 | 000 | 21 17 5 |
| 73.82480 | 5.967E-28 | 3.698963 | 4279.5121 | 000 | 22 15 8 | 000 | 22 14 9 |
| 73.83810 | 3.364E-28 | 4.465928 | 4106.7292 | 000 | 21 17 5 | 000 | 21 16 6 |
| 73.98890 | 9.021E-29 | 13.004442 | 4429.6190 | 000 | 22 17 5 | 000 | 22 16 6 |
| 74.04080 | 3.716E-28 | -3.182941 | 4353.3369 | 000 | 22 16 7 | 000 | 22 15 8 |
| 75.52250 | 1.099E-28 | 39.847444 | 4580.0103 | 000 | 22 19 4 | 000 | 22 18 5 |
| 75.86450 | 2.830E-28 | 13.145652 | 4427.3777 | 000 | 22 17 6 | 000 | 22 16 7 |
| 76.06710 | 2.141E-28 | 4.761906 | 4518.6751 | 000 | 23 13 10 | 000 | 23 12 11 |
| 76.19660 | 2.850E-29 | -6.404271 | 4683.0727 | 000 | 23 16 8 | 000 | 23 15 9 |
| 76.45210 | 4.802E-29 | -9.132908 | 4503.6079 | 000 | 22 18 4 | 000 | 22 17 5 |
| 76.76810 | 1.450E-28 | -9.198275 | 4503.2422 | 000 | 22 18 5 | 000 | 22 17 6 |
| 77.98330 | 5.065E-29 | 9.155393 | 4605.0894 | 000 | 23 15 9 | 000 | 23 14 10 |
| 78.26270 | 3.282E-29 | -13.014230 | 4837.8931 | 000 | 23 18 5 | 000 | 23 17 6 |
| 79.24770 | 1.819E-29 | -10.308079 | 5022.7903 | 000 | 24 16 9 | 000 | 24 15 10 |
| 83.35510 | 3.687E-29 | 16.922687 | 4939.4352 | 000 | 24 15 10 | 000 | 24 14 11 |
| 84.08590 | 7.657E-29 | -2.549055 | 4521.0035 | 000 | 23 14 10 | 000 | 23 13 11 |
| 85.40940 | 1.578E-27 | 1.112189 | 4116.4738 | 000 | 22 13 10 | 000 | 22 12 11 |
| 91.23680 | 3.990E-28 | -1.288501 | 4427.4383 | 000 | 23 12 11 | 000 | 23 11 12 |
| 91.27050 | 5.523E-29 | -5.152292 | 4848.1647 | 000 | 24 14 11 | 000 | 24 13 12 |
| 93.16690 | 1.409E-28 | 3.489006 | 4427.8366 | 000 | 23 13 11 | 000 | 23 12 12 |
| 100.20460 | 3.366E-29 | -3.007418 | 4746.9425 | 000 | 24 12 12 | 000 | 24 11 13 |
| 101.06070 | 1.090E-28 | 7.966880 | 4747.1040 | 000 | 24 13 12 | 000 | 24 12 13 |
| 102.95310 | 7.601E-28 | 1.821074 | 4324.4852 | 000 | 23 11 12 | 000 | 23 10 13 |
| 111.13720 | 6.617E-29 | 4.557913 | 4635.8053 | 000 | 24 11 13 | 000 | 24 10 14 |
| 111.27710 | 1.921E-28 | -1.610988 | 4635.8269 | 000 | 24 12 13 | 000 | 24 11 14 |
| 121.47370 | 3.764E-28 | 2.200825 | 4514.3532 | 000 | 24 11 14 | 000 | 24 10 15 |
| 129.24410 | 9.244E-29 | -1.525408 | 4825.8385 | 000 | 25 10 15 | 000 | 25 9 16 |
| 135.10280 | 1.463E-28 | 3.234916 | 4045.5692 | 000 | 21 17 4 | 000 | 20 20 1 |
| 139.42100 | 1.904E-28 | 2.192439 | 4686.4175 | 000 | 25 9 16 | 000 | 25 8 17 |
| 157.38230 | 9.980E-29 | 1.911638 | 4840.9796 | 000 | 26 9 18 | 000 | 26 8 19 |
| 243.34910 | 2.250E-28 | 4.106277 | 3937.3229 | 000 | 21 17 4 | 000 | 20 18 3 |
| 245.85210 | 6.097E-28 | 3.986492 | 4107.4848 | 000 | 22 15 8 | 000 | 21 16 5 |
| 246.70570 | 5.272E-28 | -3.635878 | 4180.6720 | 000 | 22 16 7 | 000 | 21 17 4 |
| 257.02790 | 2.702E-28 | 3.889352 | 4106.7292 | 000 | 22 15 7 | 000 | 21 16 6 |
| 266.95930 | 1.161E-27 | 44.126709 | 4388.5735 | 000 | 22 19 4 | 000 | 21 20 1 |
| 266.96520 | 3.869E-28 | 44.123040 | 4388.5734 | 000 | 22 19 3 | 000 | 21 20 2 |

| Wn, cm⁻¹ This work | Intensity Cm/mol | Wn, cm⁻¹ HIT - TW | $E_{lower}$, cm⁻¹ | VIB | J $K_a$ $K_c$ Upper | VIB | J $K_a$ $K_c$ Lower |
|---|---|---|---|---|---|---|---|
| 278.52130 | 1.028E-27 | 18.193729 | 4045.5692 | 000 | 21 19  3 | 000 | 20 20  0 |
| 278.52250 | 3.084E-27 | 18.193324 | 4045.5692 | 000 | 21 19  2 | 000 | 20 20  1 |
| 297.20330 | 2.806E-27 | 1.735027  | 3963.5148 | 000 | 22 13  9 | 000 | 21 14  8 |
| 299.58560 | 1.007E-27 | -1.877097 | 4353.3369 | 000 | 23 14  9 | 000 | 22 15  8 |
| 310.96450 | 1.053E-26 | 1.783344  | 4116.4738 | 000 | 23 11 12 | 000 | 22 12 11 |
| 311.34170 | 4.756E-27 | 2.225279  | 4324.4852 | 000 | 24 11 14 | 000 | 23 10 13 |
| 311.48530 | 2.243E-27 | 2.229811  | 4514.3532 | 000 | 25  9 16 | 000 | 24 10 15 |
| 311.94440 | 1.128E-27 | 1.849893  | 4686.4175 | 000 | 26  9 18 | 000 | 25  8 17 |
| 312.57970 | 6.071E-28 | 1.219128  | 4840.9796 | 000 | 27  7 20 | 000 | 26  8 19 |
| 315.23010 | 2.576E-27 | 5.068935  | 4279.5121 | 000 | 23 13 10 | 000 | 22 14  9 |
| 319.10590 | 8.695E-28 | 4.591774  | 4427.8366 | 000 | 24 11 13 | 000 | 23 12 12 |
| 319.25570 | 1.161E-27 | -1.496422 | 4635.8269 | 000 | 25 10 15 | 000 | 24 11 14 |
| 319.28560 | 3.869E-28 | 4.967560  | 4635.8053 | 000 | 25 11 15 | 000 | 24 10 14 |
| 319.57350 | 5.398E-28 | -1.526056 | 4825.8385 | 000 | 26 10 17 | 000 | 25  9 16 |
| 319.66570 | 2.667E-27 | -1.206783 | 4427.4383 | 000 | 24 12 13 | 000 | 23 11 12 |
| 324.22860 | 2.135E-27 | 3.344932  | 4196.7749 | 000 | 23 13 11 | 000 | 22 12 10 |
| 326.14360 | 4.914E-28 | -1.904650 | 4521.0035 | 000 | 24 12 12 | 000 | 23 13 11 |
| 327.03100 | 6.079E-28 | 9.361014  | 4747.1040 | 000 | 25 11 14 | 000 | 24 12 13 |
| 329.48960 | 1.475E-27 | 8.048597  | 4518.6751 | 000 | 24 13 12 | 000 | 23 12 11 |
| 344.69300 | 9.214E-28 | -1.865600 | 4594.7422 | 000 | 24 14 11 | 000 | 23 13 10 |

*⁾ Only transitions with $\Delta=|\nu_{HIT}-\nu_{TW}| \geq 1$ cm⁻¹ are included.

**Table 3.** Comparison of the presently calculated using GF approach (010)-(010) pure rotational transition wavenumbers with those from HITRAN 2016 database for $H_2^{32}S$ molecule .[*)]

| Wn, cm$^{-1}$ This work | Intensity Cm/mol | Wn, cm$^{-1}$ HIT - TW | E$_{lower}$, cm$^{-1}$ | VIB | J K$_a$ K$_c$ Upper | VIB | J K$_a$ K$_c$ Lower |
|---|---|---|---|---|---|---|---|
| 41.87520 | 7.048E-31 | 5.256199 | 3720.5943 | 010 | 18  6 12 | 010 | 17  9  9 |
| 42.39200 | 2.155E-30 | 4.818959 | 3720.0792 | 010 | 18  7 12 | 010 | 17  8  9 |
| 51.61560 | 2.029E-29 | 2.925450 | 3481.5722 | 010 | 17  6 11 | 010 | 16  9  8 |
| 52.19330 | 1.948E-28 | -3.501124 | 4123.3399 | 010 | 18 11  7 | 010 | 18 10  8 |
| 53.03120 | 7.204E-30 | 2.385056 | 3480.1619 | 010 | 17  7 11 | 010 | 16  8  8 |
| 53.33150 | 3.420E-29 | 1.378892 | 3369.4762 | 010 | 17  5 12 | 010 | 18  4 15 |
| 53.33180 | 1.140E-29 | 1.379523 | 3369.4762 | 010 | 17  6 12 | 010 | 18  3 15 |
| 55.08190 | 6.184E-32 | -3.073612 | 3533.1931 | 010 | 16 10  6 | 010 | 17  7 11 |
| 57.05730 | 3.435E-31 | 7.244876 | 3997.3381 | 010 | 18 10  9 | 010 | 17 13  4 |
| 57.37690 | 1.467E-27 | 2.101508 | 3939.9612 | 010 | 17 13  4 | 010 | 17 12  5 |
| 58.68660 | 7.757E-32 | -1.797756 | 3312.5656 | 010 | 15 11  5 | 010 | 16  6 10 |
| 59.43230 | 4.834E-31 | -1.243890 | 3491.2620 | 010 | 16 10  7 | 010 | 15 13  2 |
| 60.39700 | 1.125E-31 | -10.959077 | 3870.8149 | 010 | 17 12  6 | 010 | 18  7 11 |
| 60.89330 | 4.051E-27 | 1.019228 | 3430.2842 | 010 | 15 13  3 | 010 | 15 12  4 |
| 61.25450 | 4.259E-29 | 1.391617 | 3251.3111 | 010 | 16  6 10 | 010 | 15  9  7 |
| 61.51670 | 2.344E-31 | -1.461883 | 3551.8042 | 010 | 16 11  6 | 010 | 15 14  1 |
| 62.96820 | 1.165E-29 | 2.358405 | 3579.9717 | 010 | 18  5 13 | 010 | 19  4 16 |
| 62.96830 | 3.497E-29 | 2.358671 | 3579.9717 | 010 | 18  6 13 | 010 | 19  3 16 |
| 63.70070 | 9.988E-28 | 2.800296 | 3737.6323 | 010 | 16 14  2 | 010 | 16 13  3 |
| 64.07650 | 3.025E-27 | 2.818147 | 3737.2169 | 010 | 16 14  3 | 010 | 16 13  4 |
| 64.54440 | 5.987E-28 | 2.095163 | 3931.2119 | 010 | 17 13  5 | 010 | 17 12  6 |
| 68.76640 | 1.362E-27 | -1.683496 | 3798.0774 | 010 | 17 11  7 | 010 | 17 10  8 |
| 69.18270 | 8.916E-28 | -1.575520 | 4130.6283 | 010 | 18 12  7 | 010 | 18 11  8 |
| 72.73750 | 1.138E-30 | 8.927057 | 3798.0774 | 010 | 18  7 11 | 010 | 17 10  8 |
| 76.03800 | 2.322E-31 | -1.051570 | 3551.7984 | 010 | 16 11  5 | 010 | 15 14  2 |
| 76.13770 | 3.929E-30 | 7.215534 | 3794.6968 | 010 | 18  8 11 | 010 | 17  9  8 |
| 76.23290 | 1.439E-27 | -3.263290 | 4054.3954 | 010 | 18 11  8 | 010 | 18 10  9 |
| 77.48310 | 2.311E-27 | -1.568694 | 3720.5943 | 010 | 17 10  8 | 010 | 17  9  9 |
| 80.13310 | 7.069E-31 | -4.460848 | 3533.1878 | 010 | 16 11  6 | 010 | 17  6 11 |
| 81.68180 | 5.039E-29 | 4.304797 | 3550.6943 | 010 | 17  7 10 | 010 | 16 10  7 |
| 82.08480 | 5.151E-31 | -20.994947 | 4117.7262 | 010 | 18 12  7 | 010 | 19  7 12 |
| 84.48940 | 1.563E-28 | -5.954743 | 4319.1832 | 010 | 19 11  9 | 010 | 19 10 10 |
| 86.29250 | 2.576E-27 | -1.922226 | 3968.1029 | 010 | 18 10  9 | 010 | 18  9 10 |
| 89.32240 | 1.036E-28 | 1.699229 | 3313.0649 | 010 | 16  7  9 | 010 | 15 10  6 |
| 89.52570 | 2.320E-29 | 2.878604 | 3542.9093 | 010 | 17  8 10 | 010 | 16  9  7 |
| 89.76080 | 9.232E-31 | -5.847619 | 3762.4712 | 010 | 17 10  7 | 010 | 18  7 12 |
| 93.28920 | 1.384E-30 | 21.410597 | 4130.6283 | 010 | 19  8 11 | 010 | 18 11  8 |
| 93.80470 | 4.966E-31 | -2.435828 | 3801.2934 | 010 | 17 11  6 | 010 | 16 14  3 |
| 94.74240 | 8.980E-28 | -1.336343 | 4223.9175 | 010 | 19  9 10 | 010 | 19  8 11 |
| 95.20050 | 2.961E-28 | -2.007462 | 4223.9827 | 010 | 19 10 10 | 010 | 19  9 11 |
| 96.28120 | 2.086E-28 | 1.136105 | 3415.7675 | 010 | 18  5 14 | 010 | 19  2 17 |
| 96.28120 | 6.955E-29 | 1.136261 | 3415.7675 | 010 | 18  4 14 | 010 | 19  3 17 |
| 101.07510 | 1.608E-30 | 11.053448 | 3866.8438 | 010 | 18  8 10 | 010 | 17 11  7 |
| 101.59600 | 3.678E-31 | -1.546939 | 3211.4689 | 010 | 15 10  6 | 010 | 16  5 11 |
| 104.37430 | 4.926E-31 | -8.508389 | 3762.4695 | 010 | 17 11  7 | 010 | 18  6 12 |
| 106.19130 | 1.729E-27 | 1.991170 | 4117.7262 | 010 | 19  8 11 | 010 | 19  7 12 |
| 106.25000 | 5.758E-28 | 1.797467 | 4117.7327 | 010 | 19  9 11 | 010 | 19  8 12 |
| 106.75830 | 8.643E-29 | 4.706817 | 3613.3209 | 010 | 17  8  9 | 010 | 16 11  6 |
| 108.34540 | 2.991E-27 | 2.102164 | 3762.4695 | 010 | 18  7 11 | 010 | 18  6 12 |
| 108.36330 | 8.972E-27 | 2.103314 | 3762.4712 | 010 | 18  8 11 | 010 | 18  7 12 |
| 108.90970 | 1.370E-28 | 1.529058 | 3371.2522 | 010 | 16  8  8 | 010 | 15 11  5 |
| 110.38010 | 4.096E-26 | 1.058887 | 3422.8077 | 010 | 17  6 11 | 010 | 17  5 12 |

| Wn, cm$^{-1}$ | Intensity | Wn, cm$^{-1}$ | E$_{lower}$, cm$^{-1}$ | VIB | J K$_a$ K$_c$ | VIB | J K$_a$ K$_c$ |
|---|---|---|---|---|---|---|---|
| This work | Cm/mol | HIT - TW | | | Upper | | Lower |
| 110.38510 | 1.365E-26 | 1.059470 | 3422.8080 | 010 | 17  7 11 | 010 | 17  6 12 |
| 112.62510 | 5.609E-29 | 1.166412 | 3430.2842 | 010 | 16  9  7 | 010 | 15 12  4 |
| 115.01510 | 1.121E-29 | 2.393971 | 3737.2169 | 010 | 17 10  7 | 010 | 16 13  4 |
| 115.87090 | 1.098E-29 | 8.156679 | 3852.2320 | 010 | 18  9 10 | 010 | 17 10  7 |
| 117.33620 | 3.391E-27 | 5.522567 | 4000.3900 | 010 | 19  7 12 | 010 | 19  6 13 |
| 117.34210 | 1.131E-27 | 5.625303 | 4000.3906 | 010 | 19  8 12 | 010 | 19  7 13 |
| 118.56000 | 5.093E-31 | -1.484741 | 3312.5826 | 010 | 15 12  3 | 010 | 16  7 10 |
| 119.52960 | 5.615E-27 | 3.104608 | 3642.9399 | 010 | 18  6 12 | 010 | 18  5 13 |
| 119.53120 | 1.685E-26 | 3.106681 | 3642.9400 | 010 | 18  7 12 | 010 | 18  6 13 |
| 120.10130 | 6.171E-31 | -1.419930 | 3422.8080 | 010 | 16  9  7 | 010 | 17  6 12 |
| 120.26050 | 6.781E-29 | 4.049140 | 3674.4363 | 010 | 17  9  8 | 010 | 16 12  5 |
| 121.66240 | 7.485E-26 | 1.146262 | 3301.1453 | 010 | 17  5 12 | 010 | 17  4 13 |
| 121.66270 | 2.495E-26 | 1.146891 | 3301.1453 | 010 | 17  6 12 | 010 | 17  5 13 |
| 121.82700 | 1.590E-30 | 10.900042 | 3931.2119 | 010 | 18  9  9 | 010 | 17 12  6 |
| 126.50360 | 3.708E-31 | -8.961356 | 3870.8345 | 010 | 17 13  4 | 010 | 18  8 11 |
| 127.88660 | 2.308E-30 | -2.801562 | 3422.8077 | 010 | 16 10  7 | 010 | 17  5 12 |
| 128.55370 | 6.594E-27 | 7.117474 | 3871.8363 | 010 | 19  6 13 | 010 | 19  5 14 |
| 128.55420 | 2.198E-27 | 7.085310 | 3871.8364 | 010 | 19  7 13 | 010 | 19  6 14 |
| 130.23830 | 2.189E-30 | -13.896860 | 4000.3900 | 010 | 18 11  8 | 010 | 19  6 13 |
| 130.89120 | 1.049E-26 | 2.809550 | 3512.0487 | 010 | 18  5 13 | 010 | 18  4 14 |
| 130.89130 | 3.147E-26 | 2.809425 | 3512.0487 | 010 | 18  6 13 | 010 | 18  5 14 |
| 132.31930 | 9.645E-29 | 2.879252 | 3588.2750 | 010 | 17  9  9 | 010 | 16 10  6 |
| 134.63100 | 5.193E-28 | -1.540521 | 3542.9093 | 010 | 16 12  4 | 010 | 16  9  7 |
| 137.45630 | 7.925E-28 | 13.463488 | 4109.4637 | 010 | 20  6 14 | 010 | 20  5 15 |
| 137.45650 | 2.377E-27 | 13.398913 | 4109.4637 | 010 | 20  7 14 | 010 | 20  6 15 |
| 140.01410 | 1.257E-26 | 5.938003 | 3731.8222 | 010 | 19  5 14 | 010 | 19  4 15 |
| 140.01420 | 4.189E-27 | 5.946491 | 3731.8222 | 010 | 19  6 14 | 010 | 19  5 15 |
| 141.19730 | 2.818E-28 | 2.698775 | 3438.7744 | 010 | 19  4 16 | 010 | 20  1 19 |
| 141.19730 | 8.454E-28 | 2.698214 | 3438.7744 | 010 | 19  3 16 | 010 | 20  2 19 |
| 142.57250 | 1.938E-26 | 1.586673 | 3369.4762 | 010 | 18  4 14 | 010 | 18  3 15 |
| 142.57250 | 5.815E-26 | 1.586601 | 3369.4762 | 010 | 18  5 14 | 010 | 18  4 15 |
| 144.34720 | 2.557E-31 | -4.019921 | 3533.1931 | 010 | 16 12  4 | 010 | 17  7 11 |
| 145.10610 | 3.030E-28 | -1.010422 | 3852.2320 | 010 | 17 13  4 | 010 | 17 10  7 |
| 145.26440 | 6.226E-28 | -3.847330 | 3794.6968 | 010 | 17 12  5 | 010 | 17  9  8 |
| 147.67450 | 5.864E-28 | -1.668324 | 3480.1619 | 010 | 16 11  5 | 010 | 16  8  8 |
| 151.75680 | 3.743E-30 | -2.005539 | 3642.9400 | 010 | 17  9  8 | 010 | 18  6 13 |
| 151.85050 | 2.347E-26 | 3.430278 | 3579.9717 | 010 | 19  4 15 | 010 | 19  3 16 |
| 151.85050 | 7.823E-27 | 3.427918 | 3579.9717 | 010 | 19  5 15 | 010 | 19  4 16 |
| 155.13750 | 1.312E-30 | -3.720284 | 3642.9399 | 010 | 17 10  8 | 010 | 18  5 13 |
| 159.29730 | 6.792E-29 | 8.246105 | 3895.0981 | 010 | 18 10  9 | 010 | 17 11  6 |
| 164.20420 | 1.428E-26 | 1.587406 | 3415.7675 | 010 | 19  4 16 | 010 | 19  3 17 |
| 164.20420 | 4.282E-26 | 1.586859 | 3415.7675 | 010 | 19  3 16 | 010 | 19  2 17 |
| 165.55210 | 1.028E-30 | -1.418050 | 3211.4701 | 010 | 15 11  4 | 010 | 16  6 11 |
| 166.12100 | 6.700E-29 | -7.711310 | 4053.0389 | 010 | 18 12  6 | 010 | 18  9  9 |
| 170.24100 | 4.427E-28 | 2.290327 | 3627.8364 | 010 | 17 10  8 | 010 | 16 11  5 |
| 173.49660 | 1.487E-29 | 3.700608 | 3627.8364 | 010 | 16 14  2 | 010 | 16 11  5 |
| 175.01890 | 4.447E-28 | -3.040313 | 3720.0792 | 010 | 17 11  6 | 010 | 17  8  9 |
| 177.25280 | 2.437E-26 | 1.054645 | 3238.5147 | 010 | 19  3 17 | 010 | 19  2 18 |
| 177.25280 | 7.311E-26 | 1.054742 | 3238.5147 | 010 | 19  2 17 | 010 | 19  1 18 |
| 177.49000 | 1.017E-30 | -8.959550 | 3762.4712 | 010 | 17 12  5 | 010 | 18  7 12 |
| 178.11260 | 8.713E-29 | 1.263912 | 3313.0649 | 010 | 15 13  3 | 010 | 15 10  6 |
| 178.43250 | 1.823E-27 | 1.054464 | 3237.3350 | 010 | 19  2 17 | 010 | 20  1 20 |
| 178.43250 | 6.077E-28 | 1.054370 | 3237.3350 | 010 | 19  3 17 | 010 | 20  0 20 |
| 181.20250 | 1.727E-30 | -1.868267 | 3871.8364 | 010 | 18  9  9 | 010 | 19  6 14 |
| 182.55910 | 4.976E-30 | -3.516096 | 3871.8363 | 010 | 18 10  9 | 010 | 19  5 14 |
| 183.17430 | 5.095E-27 | -1.479874 | 3491.2620 | 010 | 16 12  5 | 010 | 15 13  2 |
| 186.36280 | 1.848E-27 | -1.393337 | 3491.1775 | 010 | 16 12  4 | 010 | 15 13  3 |

| Wn, cm$^{-1}$ | Intensity | Wn, cm$^{-1}$ | E$_{lower}$, cm$^{-1}$ | VIB | J K$_a$ K$_c$ | VIB | J K$_a$ K$_c$ |
|---|---|---|---|---|---|---|---|
| This work | Cm/mol | HIT - TW | | Upper | | Lower | |
| 186.99780 | 2.805E-26 | 4.421033 | 3438.7744 | 010 | 20  3 18 | 010 | 20  2 19 |
| 186.99780 | 9.349E-27 | 4.421684 | 3438.7744 | 010 | 20  2 18 | 010 | 20  1 19 |
| 187.97250 | 2.453E-29 | 4.102332 | 3613.3209 | 010 | 16 14  3 | 010 | 16 11  6 |
| 189.33540 | 1.596E-27 | 3.941406 | 3436.4368 | 010 | 20  3 18 | 010 | 21  0 21 |
| 189.33540 | 5.321E-28 | 3.941949 | 3436.4368 | 010 | 20  2 18 | 010 | 21  1 21 |
| 190.66710 | 2.927E-28 | 6.083093 | 3939.9612 | 010 | 18 11  8 | 010 | 17 12  5 |
| 192.86410 | 4.617E-28 | -1.170982 | 3481.5722 | 010 | 16 12  5 | 010 | 16  9  8 |
| 194.42330 | 3.768E-28 | -1.346770 | 3801.3330 | 010 | 17 13  5 | 010 | 16 14  2 |
| 194.83190 | 1.412E-27 | 3.117200 | 3606.4615 | 010 | 16 14  3 | 010 | 15 15  0 |
| 194.87180 | 4.714E-28 | 3.124228 | 3606.4612 | 010 | 16 14  2 | 010 | 15 15  1 |
| 196.04470 | 1.180E-27 | -1.434598 | 3801.2934 | 010 | 17 13  4 | 010 | 16 14  3 |
| 198.27200 | 3.921E-28 | 3.020603 | 3863.9079 | 010 | 17 14  4 | 010 | 16 15  1 |
| 198.47230 | 1.182E-27 | 3.022034 | 3863.9052 | 010 | 17 14  3 | 010 | 16 15  2 |
| 202.47290 | 3.388E-28 | 2.406065 | 3997.3381 | 010 | 18 12  7 | 010 | 17 13  4 |
| 204.35600 | 3.213E-28 | 3.340924 | 4062.3775 | 010 | 18 13  6 | 010 | 17 14  3 |
| 208.03050 | 6.921E-30 | 1.097147 | 3512.0487 | 010 | 17  8  9 | 010 | 18  5 14 |
| 210.00470 | 1.157E-25 | 3.310314 | 3415.7675 | 010 | 20  2 18 | 010 | 19  3 17 |
| 210.00470 | 3.472E-25 | 3.309678 | 3415.7675 | 010 | 20  3 18 | 010 | 19  2 17 |
| 210.49550 | 1.251E-25 | 2.037818 | 3369.4762 | 010 | 19  4 16 | 010 | 18  3 15 |
| 210.49550 | 3.752E-25 | 2.037355 | 3369.4762 | 010 | 19  3 16 | 010 | 18  4 15 |
| 210.61760 | 7.876E-29 | -3.600714 | 3720.5943 | 010 | 17 12  6 | 010 | 17  9  9 |
| 210.75620 | 6.566E-28 | -2.177310 | 3402.5647 | 010 | 16 11  6 | 010 | 16  8  9 |
| 210.90340 | 1.479E-25 | 1.354042 | 3301.1453 | 010 | 18  4 14 | 010 | 17  5 13 |
| 210.90340 | 4.438E-25 | 1.353972 | 3301.1453 | 010 | 18  5 14 | 010 | 17  4 13 |
| 211.33760 | 5.738E-25 | 1.282397 | 3211.4701 | 010 | 17  5 12 | 010 | 16  6 11 |
| 211.33910 | 1.913E-25 | 1.284087 | 3211.4689 | 010 | 17  6 12 | 010 | 16  5 11 |
| 212.40140 | 5.666E-28 | -1.017239 | 3100.6635 | 010 | 15 10  6 | 010 | 15  7  9 |
| 219.77350 | 1.849E-25 | 3.881032 | 3512.0487 | 010 | 19  4 15 | 010 | 18  5 14 |
| 219.77350 | 6.163E-26 | 3.879063 | 3512.0487 | 010 | 19  5 15 | 010 | 18  4 14 |
| 219.85590 | 3.296E-28 | -1.227039 | 3632.3761 | 010 | 17 10  7 | 010 | 17  7 10 |
| 220.13190 | 7.972E-26 | 3.016700 | 3422.8080 | 010 | 18  5 13 | 010 | 17  6 12 |
| 220.13230 | 2.392E-25 | 3.017134 | 3422.8077 | 010 | 18  6 13 | 010 | 17  5 12 |
| 220.60520 | 3.354E-25 | 2.119548 | 3312.5826 | 010 | 17  6 11 | 010 | 16  7 10 |
| 220.62750 | 1.118E-25 | 2.116358 | 3312.5656 | 010 | 17  7 11 | 010 | 16  6 10 |
| 220.66180 | 7.191E-27 | 1.302089 | 3674.4363 | 010 | 17 11  6 | 010 | 16 12  5 |
| 228.85080 | 1.009E-25 | 1.122918 | 3251.3111 | 010 | 16  8  8 | 010 | 15  9  7 |
| 228.89630 | 9.490E-26 | 7.009610 | 3642.9400 | 010 | 19  5 14 | 010 | 18  6 13 |
| 228.89650 | 3.163E-26 | 7.016004 | 3642.9399 | 010 | 19  6 14 | 010 | 18  5 13 |
| 229.27640 | 4.454E-26 | 5.061838 | 3533.1931 | 010 | 18  6 12 | 010 | 17  7 11 |
| 229.28340 | 1.337E-25 | 5.064928 | 3533.1878 | 010 | 18  7 12 | 010 | 17  6 11 |
| 229.81140 | 2.018E-25 | 2.727885 | 3402.5647 | 010 | 17  7 10 | 010 | 16  8  9 |
| 229.84440 | 4.998E-26 | 1.411096 | 3313.0649 | 010 | 16  9  7 | 010 | 15 10  6 |
| 230.04770 | 6.733E-26 | 2.590471 | 3402.3873 | 010 | 17  8 10 | 010 | 16  7  9 |
| 231.70810 | 1.509E-28 | -6.761036 | 3968.1029 | 010 | 18 12  7 | 010 | 18  9 10 |
| 234.40880 | 1.044E-28 | -3.845754 | 3632.4350 | 010 | 17 11  7 | 010 | 17  8 10 |
| 236.09670 | 2.668E-30 | 4.580155 | 3731.8222 | 010 | 18  8 10 | 010 | 19  5 15 |
| 236.28070 | 8.122E-30 | 4.344133 | 3731.8222 | 010 | 18  9 10 | 010 | 19  4 15 |
| 237.91880 | 5.030E-26 | 11.020403 | 3762.4712 | 010 | 19  6 13 | 010 | 18  7 12 |
| 237.92110 | 1.677E-26 | 10.996707 | 3762.4695 | 010 | 19  7 13 | 010 | 18  6 12 |
| 238.11170 | 6.795E-28 | -1.740902 | 3312.5826 | 010 | 16 10  7 | 010 | 16  7 10 |
| 238.37990 | 2.557E-26 | 6.764799 | 3632.4350 | 010 | 18  7 11 | 010 | 17  8 10 |
| 238.45840 | 7.673E-26 | 6.723894 | 3632.3761 | 010 | 18  8 11 | 010 | 17  7 10 |
| 238.50700 | 1.231E-25 | 3.171420 | 3481.5722 | 010 | 17  8  9 | 010 | 16  9  8 |
| 238.91110 | 2.740E-26 | 3.678156 | 3613.3209 | 010 | 17 10  7 | 010 | 16 11  6 |
| 240.43240 | 4.151E-26 | 2.190696 | 3480.1619 | 010 | 17  9  9 | 010 | 16  8  8 |
| 244.00250 | 6.995E-26 | 3.813157 | 3550.6943 | 010 | 17  9  8 | 010 | 16 10  7 |
| 246.89170 | 2.716E-26 | 14.439656 | 3870.8345 | 010 | 19  7 12 | 010 | 18  8 11 |

| Wn, cm$^{-1}$ | Intensity | Wn, cm$^{-1}$ | E$_{lower}$, cm$^{-1}$ | VIB | J K$_a$ K$_c$ | VIB | J K$_a$ K$_c$ |
|---|---|---|---|---|---|---|---|
| This work | Cm/mol | HIT - TW | | | Upper | | Lower |
| 246.91780 | 9.055E-27 | 14.519846 | 3870.8149 | 010 | 19  8 12 | 010 | 18  7 11 |
| 247.32460 | 1.497E-26 | 7.801258 | 3720.5943 | 010 | 18  8 10 | 010 | 17  9  9 |
| 248.02370 | 4.480E-26 | 7.128000 | 3720.0792 | 010 | 18  9 10 | 010 | 17  8  9 |
| 252.15810 | 1.050E-30 | -4.752590 | 3642.9400 | 010 | 17 11  6 | 010 | 18  6 13 |
| 254.96150 | 8.719E-27 | 8.868022 | 3798.0774 | 010 | 18  9  9 | 010 | 17 10  8 |
| 255.81460 | 1.472E-26 | 16.225081 | 3968.1029 | 010 | 19  8 11 | 010 | 18  9 10 |
| 256.06380 | 4.913E-27 | 15.874417 | 3967.9189 | 010 | 19  9 11 | 010 | 18  8 10 |
| 258.42370 | 1.400E-30 | 2.126627 | 3542.9093 | 010 | 16 14  2 | 010 | 16  9  7 |
| 259.69860 | 2.667E-26 | 5.499054 | 3794.6968 | 010 | 18 10  9 | 010 | 17  9  8 |
| 259.79380 | 2.000E-28 | -4.979771 | 3870.8345 | 010 | 18 11  8 | 010 | 18  8 11 |
| 262.89990 | 7.841E-30 | 2.882127 | 3369.4762 | 010 | 17  7 10 | 010 | 18  4 15 |
| 262.95880 | 2.614E-30 | 2.838196 | 3369.4762 | 010 | 17  8 10 | 010 | 18  3 15 |
| 264.88430 | 1.169E-28 | -1.763054 | 3533.1931 | 010 | 17 10  8 | 010 | 17  7 11 |
| 275.15300 | 3.665E-30 | -1.828654 | 3402.3873 | 010 | 16 12  4 | 010 | 16  7  9 |
| 277.25890 | 3.747E-30 | -2.039083 | 3720.0792 | 010 | 17 13  4 | 010 | 17  8  9 |
| 278.39630 | 1.783E-26 | 2.971163 | 3852.2320 | 010 | 18 11  8 | 010 | 17 10  7 |
| 290.56940 | 8.419E-29 | 2.043129 | 3762.4695 | 010 | 18  9  9 | 010 | 18  6 12 |
| 290.84320 | 2.661E-30 | 7.565177 | 3579.9717 | 010 | 18  7 11 | 010 | 19  4 16 |
| 290.86280 | 7.921E-30 | 7.568666 | 3579.9717 | 010 | 18  8 11 | 010 | 19  3 16 |
| 297.27150 | 3.872E-28 | 1.304856 | 3422.8077 | 010 | 17  8  9 | 010 | 17  5 12 |
| 304.71290 | 1.724E-26 | 3.407295 | 3895.0981 | 010 | 18 12  7 | 010 | 17 11  6 |
| 307.58510 | 7.970E-30 | -4.338969 | 3632.3761 | 010 | 17 12  5 | 010 | 17  7 10 |
| 310.03140 | 2.666E-25 | 2.258043 | 3491.2620 | 010 | 16 14  3 | 010 | 15 13  2 |
| 310.15550 | 8.892E-26 | 2.273811 | 3491.1775 | 010 | 16 14  2 | 010 | 15 13  3 |
| 318.21600 | 2.881E-26 | 2.320010 | 3677.5403 | 010 | 17 13  5 | 010 | 16 12  4 |
| 318.41960 | 4.979E-30 | 2.900155 | 3214.7682 | 010 | 17  6 11 | 010 | 18  3 16 |
| 318.42490 | 1.660E-30 | 2.901368 | 3214.7682 | 010 | 17  7 11 | 010 | 18  2 16 |
| 319.72120 | 1.391E-29 | 2.566934 | 3481.5722 | 010 | 16 14  3 | 010 | 16  9  8 |
| 322.90180 | 8.966E-26 | 2.303318 | 3674.4363 | 010 | 17 13  4 | 010 | 16 12  5 |
| 324.97900 | 1.084E-28 | 5.649668 | 3642.9399 | 010 | 18  8 10 | 010 | 18  5 13 |
| 325.16290 | 3.245E-28 | 5.415740 | 3642.9400 | 010 | 18  9 10 | 010 | 18  6 13 |
| 327.97470 | 1.370E-30 | -8.354906 | 3871.8363 | 010 | 18 12  7 | 010 | 19  5 14 |
| 331.23080 | 4.371E-28 | 2.649497 | 3301.1453 | 010 | 17  7 10 | 010 | 17  4 13 |
| 331.28970 | 1.456E-28 | 2.605565 | 3301.1453 | 010 | 17  8 10 | 010 | 17  5 13 |
| 344.40380 | 6.102E-26 | 1.066106 | 3550.6943 | 010 | 17 11  6 | 010 | 16 10  7 |
| 346.70370 | 3.404E-30 | 7.052211 | 3415.7675 | 010 | 18  7 12 | 010 | 19  2 17 |
| 361.85370 | 4.796E-30 | -1.976885 | 3312.5826 | 010 | 16 12  5 | 010 | 16  7 10 |
| 361.91030 | 1.847E-30 | -2.794344 | 3533.1878 | 010 | 17 11  6 | 010 | 17  6 11 |

*) Only transitions with $\Delta=|\nu_{HIT}-\nu_{TW}| \geq 1$ cm$^{-1}$ are included.